\documentclass[journal]{IEEEtran}
\usepackage{amsmath}
\usepackage{amssymb}
\usepackage{amsfonts}
\usepackage{mathtools}
\usepackage{bm}

\def\C{\mathcal{C}}
\def\D{\mathcal{D}}

\def\G{\mathcal{G}}

\def\I{\mathcal{I}}

\def\S{\mathcal{S}}
\def\T{\mathcal{T}}

\def\V{\mathcal{V}}

\def\Z{\mathcal{Z}}
\def\aa{{\bm a}}
\def\bb{{\bm b}}

\def\ee{{\bm e}}
\def\ii{{\bm i}}
\def\ll{{\bm l}}
\def\ull{\underline{\ll}}

\def\tt{{\bm t}}
\def\uu{{\bm u}}
\def\xx{{\bm x}}
\def\yy{{\bm y}}
\def\zz{{\bm z}}
\def\ss{{\bm s}}

\def\bDelta{{\bm \Delta}}
\def\bL{{\bm L}}
\def\ubL{\underline{\bL}}

\def\ZZ{\{0,1\}}
\def\RR{\mathbb{R}}

\def\IN{{\rm in}}
\def\HD{{\rm HD}}

\def\blambda{\bm \lambda}
\def\bgamma{\bm \gamma}

\def\sign{{\rm sign}}

\def\SDD{{\rm SDD}}
\def\*{\times}

\newcommand{\ad}[1]{\textcolor{red}{#1}}

\usepackage{url}
\usepackage{cite}
\usepackage{color}
\usepackage{here}
\usepackage{algorithmic}
\usepackage{algorithm}

\ifCLASSINFOpdf
  \usepackage[pdftex]{graphicx}
  \graphicspath{{../pdf/}{pdf/}}
\else
  \usepackage[dvipdfmx]{graphicx}
\fi

\usepackage{amsmath}
\usepackage[percent]{overpic}
\begin{document}
\title{Iterative Decoder of Channel-polarized Multilevel Coding for Data Center Networks}

\author{Takeshi~Kakizaki,
        Masanori~Nakamura,~\IEEEmembership{Member,~IEEE,}
        Fukutaro~hamaoka,~\IEEEmembership{Member,~IEEE,}
        Shuto~Yamamoto,~\IEEEmembership{Member,~IEEE,}
        and Etsushi~Yamazaki,~\IEEEmembership{Member,~IEEE}
\thanks{T. Kakizaki, M. Nakamura, F. Hamaoka, S. Yamamoto, Etsushi Yamazaki are NTT Network Innovation Laboratories, Nippon Telegraph and Telephone Corporation, Yokosuka-shi, Kanagawa 239-0847, Japan, e-mail: takeshi.kakizaki@ntt.com}
\thanks{Accepted for publication in: Journal of Lightwave Technology, doi: 10.1109/JLT.2025.3603040, \copyright 2025 IEEE}}

\markboth{Accepted for publication in: Journal of Lightwave Technology, doi: 10.1109/JLT.2025.3603040, \copyright 2025 IEEE}%
{Shell \MakeLowercase{\textit{et al.}}: Bare Demo of IEEEtran.cls for IEEE Journals}

\maketitle
\begin{abstract}
Data center networks (DCNs) require a low-cost, low-power optical transceiver to handle increased traffic from generative artificial intelligence, video streaming services, and more. Improving the required signal-to-noise ratio (RSNR) by digital signal processing such as forward error correction (FEC) mitigates the requirements for electrical and optical components. The optical transceivers in DCNs exploit a low-complexity soft-decision (SD) FEC, consisting of short block-length linear error-correcting codes and a low-complexity SD decoder (SDD), such as Chase decoding and ordered statistics decoding. The low-complexity SDD efficiently approaches a maximum likelihood decoding (MLD). However, the decoding performance of MLD is limited by its finite block length. In this paper, we describe the details of our proposed channel-polarized multilevel coding with iterative decoding (CP-MLC-ID). The proposed CP-MLC-ID improves the decoding performance by extending the codeword length to weakly and indirectly connect codewords via bypassed bits. The 19.5$\%$–OH CP-MLC-ID using 128-bit extended Bose–Chaudhuri–Hocquenghem (eBCH) and KP4 codes outperforms the concatenated eBCH and KP4 codes with a net coding gain of 0.25 and 0.40 dB for the same and double the number of SDDs, respectively. We also investigate the dependency of the decoding performance on the size of a bit interleaver. The performance degradation of CP-MLC-ID using an 8-bit interleaver is about 0.1 dB compared to using the large-bit interleaver. Our results indicate that even a weak connection by exclusive-OR between codewords improves the decoding performance, compared to simple concatenated codes in the DCNs.
\end{abstract}

\IEEEpeerreviewmaketitle

\section{Introduction}
\begin{figure*}[!ht]
  \centering
  \begin{overpic}[scale=.22]{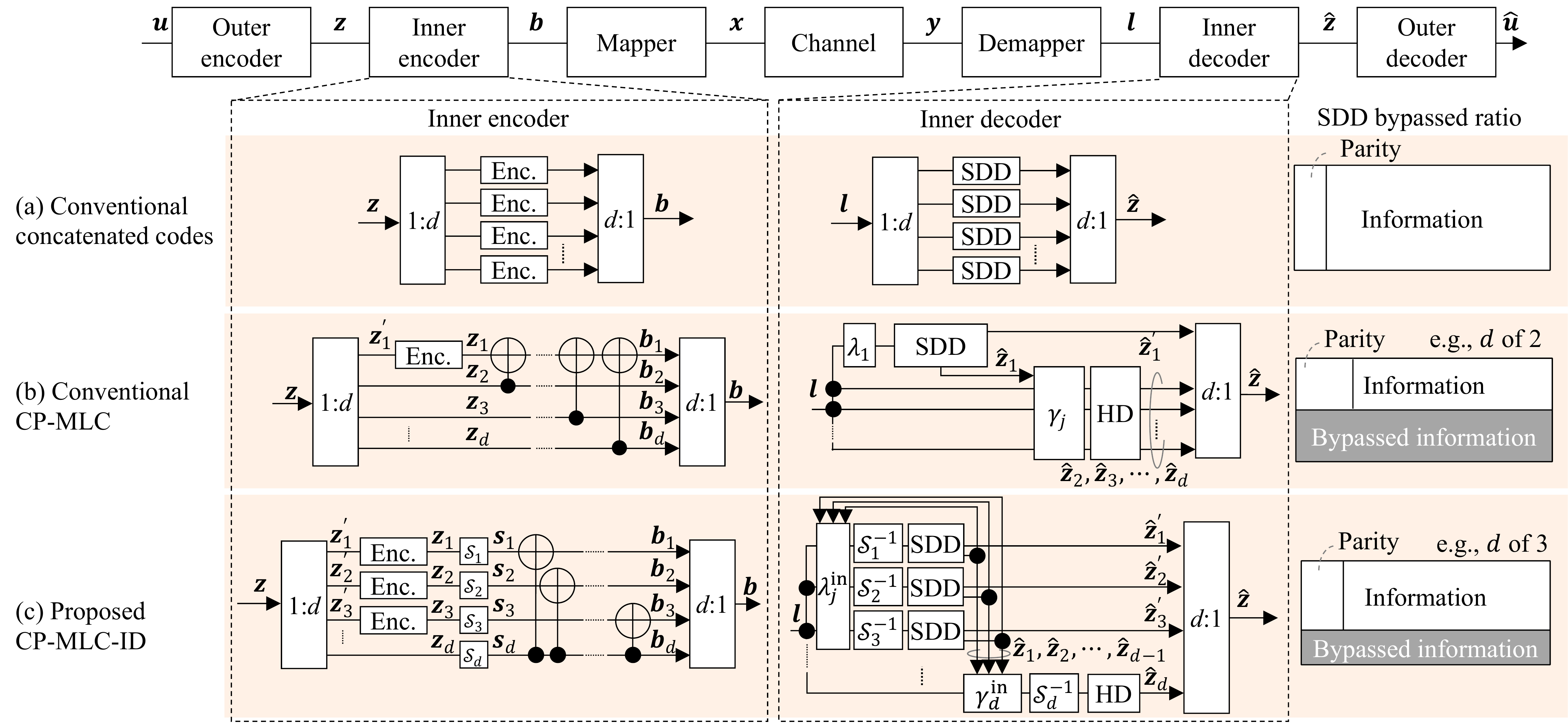}
  \end{overpic}  
  \caption{Schematic diagrams of (a) conventional concatenated codes, (b) conventional CP-MLC, and (c) proposed CP-MLC-ID.}
  \label{figure:diagram}
\end{figure*}

\IEEEPARstart{I}{nternet} traffic has been increasing rapidly to keep up with generative artificial intelligence, video streaming services, and more. Digital coherent optical communication systems have recently been deployed in intra- and inter-data center networks (DCNs) in addition to metro and core networks to achieve cost- and energy-efficient networks \cite{7937054,800lrir}. The optical transceivers in DCNs exploit a low-complexity soft-decision forward error correction (SD-FEC) scheme that consists of a pair of short block-length codes and low-complexity SD decoding (SDD) to maintain reliable communication with the low power consumption of the optical transceiver. Vendors are developing the \textit{cFEC}, which consists of the concatenated Hamming and staircase codes, for $<$120-km optical links for 400 Gbps \cite{400zr}. Novel O-band $<$10-km optical links for 800 Gbps adopt the concatenated codes, consisting of an inner Bose–Chaudhuri–Hocquenghem (BCH) and outer KP4 codes, called LR-FEC. Vendors are also considering adopting the LR-FEC for the $<$120-km optical link for a 1.6-Tbps application in the \textit{Optical Internetworking Forum} \cite{OIF_1600ZR_LR_4,OIF_1600ZR_LR_3,OIF_1600ZR_LR_2,OIF_1600ZR_LR_1}. 

The low-complexity SDD efficiently approaches the maximum likelihood decoding (MLD) by searching for codewords around the received signals. However, the decoding performance is limited by the block length due to the increasing decoding complexity of the low-complexity SDD in proportion to the minimum distance between codewords or the overhead (OH)\cite{5452208, Chase, OSD}. To improve the performance-complexity tradeoff, the iterative decoding of short block length inner and outer codes has been developed. The LR-FEC with iterative decoding provides the pre-FEC BER threshold of $1.6\times 10^{-2}$ compared to the conventional one of $1.2\times 10^{-2}$ to achieve a post-FEC BER of $10^{-15}$; however, this comes at the cost of doubling the number of iterations\cite{OIF_1600ZR_LR_1}.

As an alternative, the channel-polarized multilevel coding (CP-MLC) with low-complexity SDD and low-OH inner codes can improve the decoding complexity at the middle BER region such as the KP4 BER threshold\cite{9901044,cpmlc_ecoc2023}. The CP-MLC converts a $d$ \textit{discrete memoryless channel} $(W_1,W_2,\cdots,W_d)$ into $(V_1,V_2,\cdots,V_d)$, where we denote the unreliable bit channel $V_1$ and the reliable bit channels $V_{j}$ for each $2\leq j\leq d$, and only applies the SDD in the unreliable bit channels $V_1$\cite{cpmlc,cpmlc_jlt}. However, the CP-MLC with a high OH region cannot efficiently achieve the target BER in the middle BER region due to the error floor caused by bypassing the SDD on the reliable bit channels $V_{j}$.

To improve the decoding performance in the high-OH and middle-BER region, we proposed CP-MLC with iterative decoding (CP-MLC-ID), which uses the channel conversion $(W_1,W_2,\cdots,W_d)$ into $(U_1,U_2,\cdots,U_d)$, where we denote $d-1$ unreliable bit channels $U_j$ for each $1\leq j\leq d-1$ and a highly reliable bit channel $U_d$ instead of the channel conversion of CP-MLC, and uses iterative decoding\cite{cpmlcid,cpmlchid}. The CP-MLC-ID with extended BCH (eBCH) and KP4 codes can improve the performance-complexity tradeoffs compared to concatenated eBCH–KP4 codes and reduce the error floor in the highly reliable bit channel $U_d$.

In this paper, we explain the details of CP-MLC-ID as an extension of \cite{cpmlcid,cpmlchid}, including a discussion of interleaver design and its dependency. We show that a roughly 19.5$\%$–OH CP-MLC-ID can exceed the NCG with the same complexity and the maximum NCG by 0.25 and 0.4 dB, respectively, compared to concatenated codes. In practice, we also investigate the performance degradation of CP-MLC-ID with 128-bit eBCH codes, which is about 0.1 dB for a bit interleaver size of 8 bits. CP-MLC-ID, which exploits a weak connection by exclusive-OR (XOR) between codewords, can improve the decoding performance compared to simple concatenated codes.

Sec. \ref{sec:conventional} of this paper explains the conventional concatenated codes and CP-MLC. In Sec. \ref{sec:cpmlcid}, we describe the encoder, decoder, and interleaver design of CP-MLC-ID. In Sec. \ref{section:sim_results}, we evaluate the decoding performance and the decoding complexity of concatenated codes, CP-MLC, and CP-MLC-ID. We conclude in Sec. \ref{section:conclusion} with a brief summary.

\section{system model and conventional codes for DCNs}\label{sec:conventional}
\subsection{Concatenated codes}
This section describes the concatenated codes for the DCNs. Figure \ref{figure:diagram}(a) shows a schematic diagram of the inner encoder and decoder for the concatenated codes. First, the encoder converts the information bits $\uu\in\{0,1\}^k$ into the outer codeword $\zz\in\{0,1\}^{n_\HD}$, and the inner encoder converts the outer codeword $\zz$ into inner codewords $\bb\in\{0,1\}^n$. We assume that the modulation format is a \textit{binary phase-shift keying} (BPSK), i.e., the symbol mapper transforms $\bb$ into the symbols $\xx = 1-2\bb$. A received signal is given by $\yy\triangleq\xx+\ee^{\text{AWGN}}$, where $\ee^{\text{AWGN}}\in\RR^n$ is a vector of additive white Gaussian noise (AWGN). The output of the demapper is the log-likelihood ratio (LLR), denoted by $\ll\triangleq 2\yy/\sigma^2$.

The concatenated codes can efficiently correct large bit errors by the inner SDD and residual errors by the outer hard-decision decoder (HDD). The performance of low-complexity SDDs (e.g., Chase decoding and ordered statistics decoding (OSD)) can approach that of the MLD, which can achieve the optimal block error rate for SDD, while the MLD is not optimal for the bit error rate (BER) \cite{bervsbler}.

\subsection{Channel-polarized multilevel coding}
In this section, we explain the encoder and decoder of CP-MLC\cite{cpmlc,cpmlc_jlt}. The CP-MLC can improve the BER compared to the concatenated codes with near MLD under some BER regions\cite{cpmlc_ecoc2023}.

Figure \ref{figure:diagram}(b) shows a schematic diagram of the CP-MLC encoder and decoder. The encoder first calculates the outer codeword $\zz\triangleq (\zz_{1}',\zz_{2},\cdots,\zz_{d})$, where $\zz_{1}'\in \ZZ^{nR_{1}/d},\zz_{2},\cdots,\zz_{d}\in \ZZ^{n/d}$. The inner encoder transforms $\zz_{1}'$ into $\zz_{1}\in\ZZ^{n/d}$, where $n$ and $R_1$ denote the block length and the code rate of the inner and outer code, respectively. The encoder then converts $(\zz_{1},\zz_{2},\cdots,\zz_{d})$ into
\begin{align}
\left(\bb_{1},\bb_{2},\cdots,\bb_{d}\right)&\triangleq\left(\zz_1\oplus\zz_2\oplus\cdots\zz_{d},\zz_{2},\cdots,\zz_{d}\right),
\end{align}
where $\aa\oplus \bb$ is the XOR for each element of a vector.

On the decoder side, the decoder first calculates the estimated codeword $\hat{\zz}_{1}$ and information bits $\hat{\zz}_{1}'$ from the unreliable LLR (uLLR), as
  \begin{align}
  \lambda_1(\underline{\ll}) &\triangleq\log\frac{P_{Z_{1}}(0)}{P_{Z_{1}}(1)}+\log\frac{P_{\ubL|Z_{1}}(\ull|0)}{P_{\ubL|Z_{1}}(\ull|1)}\\
  &=l_{1}\boxplus l_{2}\boxplus \cdots\boxplus l_{d},\label{equation:SD-LLR}
  \end{align}
where we define
\begin{align}
    \ull\triangleq \left(l_1,l_2,\cdots,l_d\right)
\end{align}
and
\begin{align}
    a\boxplus b&\triangleq 2\tanh^{-1}\left(\tanh\frac{a}{2}\cdot\tanh \frac{b}{2}\right).
\end{align}

Next, the estimated bits $\zz_2,\zz_3,\cdots,\zz_d$ are calculated from $\hat{z}_j=0.5\left(1-\sign(\gamma_j(\ull,\hat{z}_1))\right)$ by using the reliable LLR (rLLR), as
  \begin{align}
    &\gamma_{j}(\ull,z_1)\triangleq\log\frac{P_{Z_{j}}(0)}{P_{Z_{j}}(1)}+\log\frac{P_{\ubL,Z_{1}|Z_{j}}(\ull,z_{1}|0)}{P_{\ubL,Z_{1}|Z_{j}}(\ull,z_{1}|1)}\\
    &=l_{j}+(-1)^{z_{1}}(l_{1}\boxplus l_{2}\boxplus\cdots\boxplus l_{j-1}\boxplus l_{j+1} \boxplus \cdots\boxplus l_{d}).\label{equation:HD-LLR}
  \end{align}
The outer decoder corrects both the residual bit error of the corrected inner information bits $\zz_1'\neq\hat{\zz}_1'$ and the bit errors of the reliable bit channels $\zz_j\neq\hat{\zz}_j$ for each $j$.

\begin{figure}[!ht]
  \centering
  \begin{overpic}[scale=0.3]{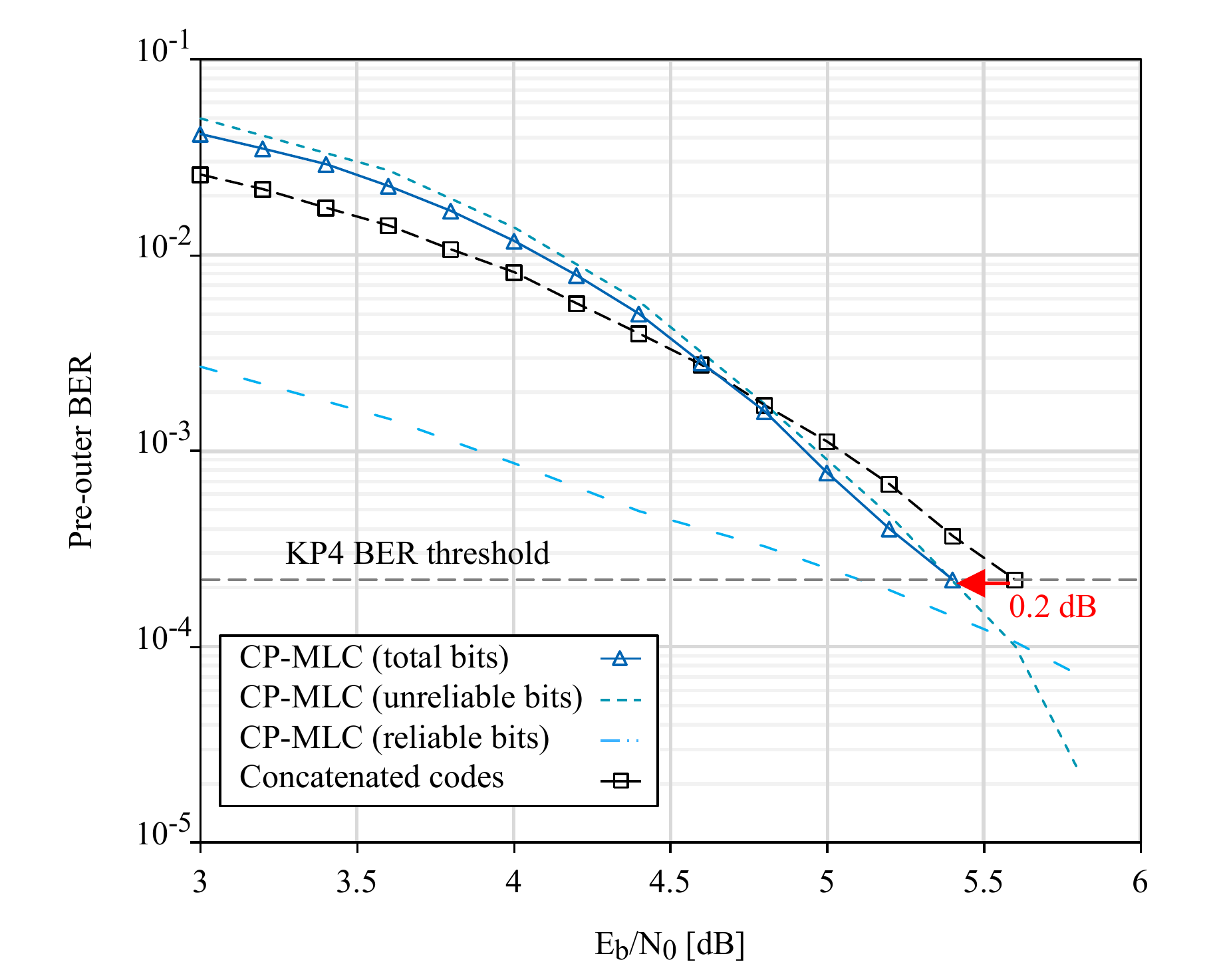}
  \end{overpic}
  \caption{Pre-outer FEC BER of concatenated codes and CP-MLC\cite{cpmlc_ecoc2023}.}
  \label{figure:CPMLC_improvement}
\end{figure}

Figure \ref{figure:CPMLC_improvement} shows the decoding performance of 12.9$\%$–OH concatenated $(128,120,4)$-eBCH–KP4 codes and CP-MLC with $(128,113,6)$-eBCH and KP4 codes with near MLD, using Chase decoding with the number of test patterns by 1024. Here, $(n,k,\mu_{\text{min}})$-eBCH codes have codeword length $n$, information length $k$, and minimum distance between codewords $\mu_{\text{min}}$. The BER of CP-MLC, shown as a solid line, is better than that of concatenated codes because high-OH eBCH codes can be exploited in unreliable bits. On the other hand, an error floor appears in the high SNR region due to the dominant error of bypassed reliable bits in the KP4 BER threshold.

\section{CP-MLC-ID}\label{sec:cpmlcid}
This section describes the encoder and the iterative decoder for CP-MLC-ID.
\subsection{Encoder}
We decompose as the information bits as $\zz \triangleq (\zz'_1,\zz'_2,\cdots,\zz'_{d-1},\zz_d)$. The inner encoder first converts each information bits $\zz'_{j}\in\{0,1\}^{nR_{j}/d}$ into $\zz_j\in\{0,1\}^{n/d}$ for each $1\leq j \leq d-1$, and $\zz_d\in \{0,1\}^{n/d}$ is bypassed. Here, the code rate $R_j$ for each inner code. The bit-interleavers $\S_1,\S_2,\cdots, \S_d$ output the bit-interleaved bits $\ss_1,\ss_2,\cdots,\ss_d\in\{0,1\}^{n/d}$, respectively. The CP-MLC-ID codeword is given by
\begin{align}
\bb\triangleq(\ss_1\oplus\ss_d,\ss_2\oplus\ss_d,\cdots,\ss_{d-1}\oplus\ss_d,\ss_d)\label{equation:encoding}.
\end{align}

\subsection{Iterative decoder}
CP-MLC-ID uses the iterative decoder to reduce the performance degradation by considering the correlation between unreliable bit channels as similar to bit-interleaved coded modulation with an iterative decoder (BICM-ID) \cite{1025496}.

On the other side, the iterative decoder with maximum iteration $I \geq 1$ first calculates the processing loop for $i=1,2,\cdots,I$ and $j=(i-1 \text{ mod } (d-1))+1$, as follows.
\begin{enumerate}
    \item 
    We calculate the uLLR
    \begin{align}
        \blambda^\IN_j&\triangleq \ll_j\boxplus\tilde{\ll}_{d,j},\label{equation:lambda}
    \end{align}
    and
    \begin{align}
    \tilde{\ll}_{d,j}&\triangleq \ll_d+\sum_{j'=1:j'\neq j}^{d-1}\ll_{d,j'}^{\text{ext}},
    \end{align}
    where $\ll_{d,j'}^{\text{ext}}={\bf 0}$ in the first iteration $i=1$, and in subsequent iterations, it is updated using Eq. \eqref{eq:extrinsic_information} based on the previous iteration.
    \item The inner decoder $\D_j$ calculates a soft-information 
    \begin{align}
    \blambda^{\text{out}}_j=\S_j\D_j(\S_j^{-1}(\blambda_j^{\IN})),
    \end{align}
    and outputs each inner codeword $\hat{\zz}_j\triangleq \S^{-1}_j(\hat{s}_j)$, where we define $\hat{s}_j\triangleq 0.5(1 - \sign(\blambda_j^{\text{out}}))$.
    \item The extrinsic information $\ll^{\text{ext}}_{d,j}$ is updated by
    \begin{align}
        \ll^{\text{ext}}_{d,j}&\triangleq \ll_j\boxplus\blambda^{\text{out}}_j\\
        &\simeq\sign(\ll_j)\sign(\blambda_j^{\text{out}})\min(|\ll_j|,|\blambda_j^{\text{out}}|)\label{eq:minsum}\\
        &\simeq\xi[i]\ll_j(-1)^{\hat{\ss}_j}\label{eq:extrinsic_information}\\
        &=\xi[i]\underbrace{\ll_j(-1)^{\ss_j}}_{\text{ Correctness}}-\xi[i]\underbrace{2\ee_j^{\ss}\cdot\ll_j(-1)^{\ss_j}}_{\text{Incorrectness }\bDelta_j},
    \end{align}
    where a bit error is defined by $\ee_j^{\ss}\triangleq\ss_j\oplus\hat{\ss}_j$ and $\xi[i]$ is the \textit{damping factor} for each iteration. With the small value of $\xi[i]$, the incorrectness $\bDelta_j$ decreases at the cost of fewer correct updates of the LLR $\ll_{d,j}^\text{ext}$. 
    The approximation in Eq. \eqref{eq:extrinsic_information} is based on the assumption that $\blambda_j^{\text{out}}$ is more reliable than $\ll_j$. This approximation reduces the complexity by considering only the most likely codeword $\hat{\ss}_j$. Equation \eqref{eq:minsum} requires soft information, which is computed using both the most likely and the second most likely codewords, similar to the Chase–Pyndiah algorithm \cite{TPC_pyndiah}. Computing this soft information involves identifying the second most likely codeword for each bit position, which increases the number of candidate codewords. An additional benefit of this approximation is that the bit interleaver operates solely on binary bits, rather than on $\blambda_j^{\text{out}}$, thereby reducing the complexity.
\end{enumerate} 

Next, bypassed bits $\hat{\zz}_d$ are calculated by 
\begin{align}
\hat{\zz}_d\triangleq\S^{-1}_d(0.5(1-\sign(\bgamma^\IN_d))),\label{equation:reliable_bits}
\end{align}
where rLLR is given by
\begin{align}
\bgamma_d^\IN\triangleq\sum_{j=1}^{d-1}\ll_{d,j}^{\text{ext}}.
\end{align}
The probabilistic behavior of iterative decoding can be approximately modeled as belief propagation on a factor graph (see Appendix \ref{appendix:factorgraph} and \cite{cpmlchid}).

\subsection{Interleaver design}
In this subsection, we describe the design of the bit interleaver $\S_j$, which mitigates the incorrectness $\bDelta_j$ in the inner codeword $\zz_j$. Consider the CP-MLC-ID with $d=3$ over a \textit{binary symmetric channel} (BSC) with the bit-flip probability $p$. We assume the damping factor $\xi[i]$ is equal to $1$. In this scenario, the error propagation induced by the XOR operation can be expressed as
\begin{align}
\ll^{\text{ext}}_{d,j'}&\simeq (-1)^{\hat{\bb}_{j'}}(-1)^{\hat{\ss}_{j'}}=
(-1)^{\bb_d \oplus \ee_{j'}^\bb\oplus\ee_{j'}^\ss}
\end{align}
and 
\begin{align}
\blambda_{j}^{\IN}=(-1)^{\bb_j\oplus\ee_j^\bb}\boxplus\left((-1)^{\bb_d\oplus \ee_d^\bb}+(-1)^{\bb_d \oplus \ee_{j'}^\bb\oplus \ee_{j'}^\ss}\right)
\end{align}
from $\ss_d=\bb_d$ and $\bb_j=\ss_j\oplus\ss_d$ in \eqref{equation:encoding}. At the bit position where $e_{j'}^\ss=1$ and $e_d^\bb\oplus e_{j'}^\bb=0$, an uLLR $\lambda_{j}^{\IN}=(-1)^{b_j\oplus e_j^b}\boxplus 0$ becomes zero, i.e., \textit{bit erasure} $\epsilon$, which occurs with the conditional erasure probability $P_{\Lambda_{j}^\IN|E_{j'}^\ss}(\epsilon|1)=(1-p)^2+p^2$. This probability is greater than that of $e_d^\bb\oplus e_{j'}^\bb=1$ and $e_{j'}^\ss=0$, which occurs with probability $P_{\Lambda_{j}^\IN|E_{j'}^\ss}(\epsilon|0)=2(1-p)p$. Thus, the decoding performance deteriorates as $|\ee_{j'}^\ss|$ increases.

\begin{figure}[!ht]
  \centering
  \begin{overpic}[scale=0.2]{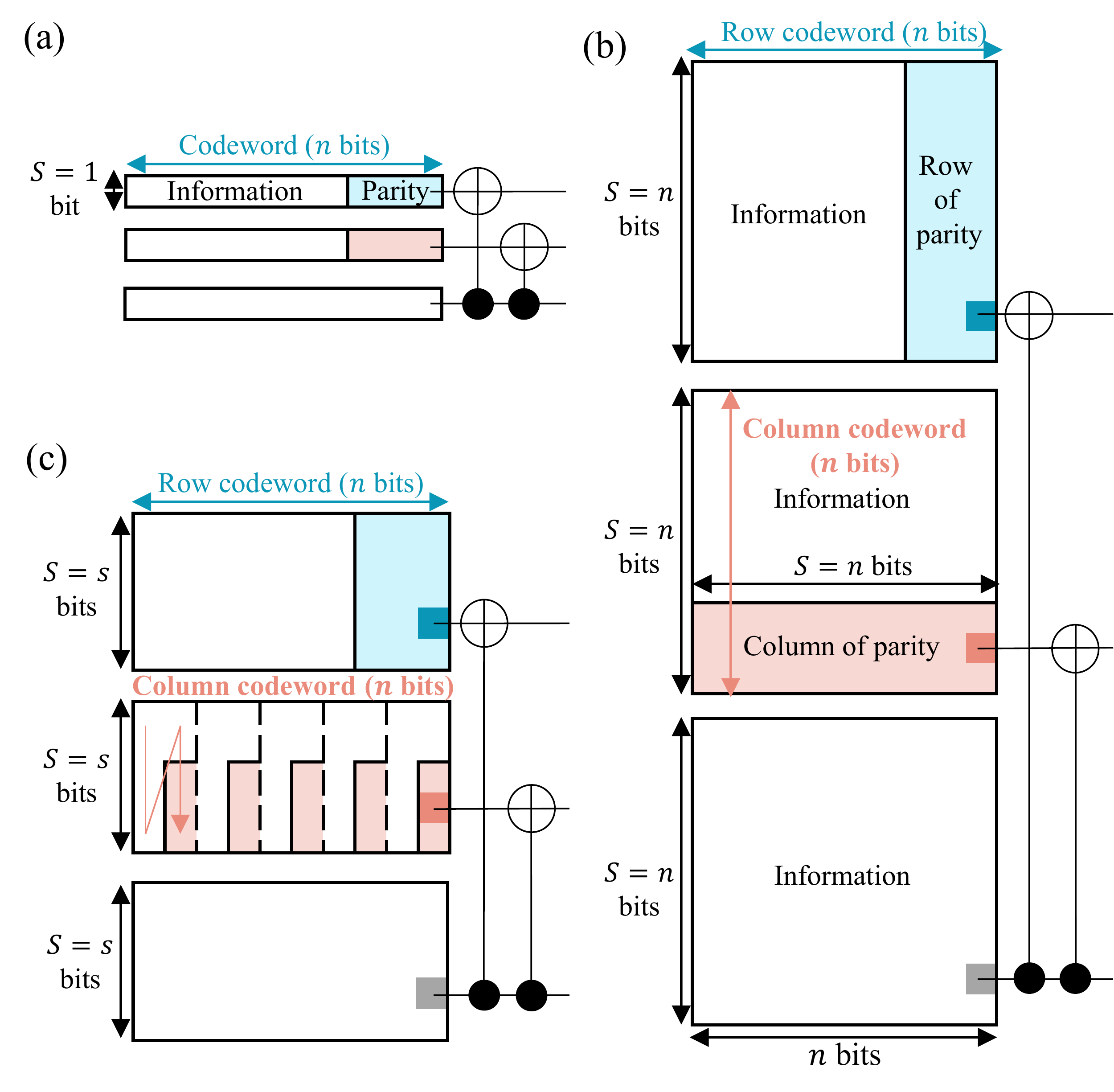}
  \end{overpic}  
  \caption{CP-MLC-ID with $d=3$ using bit interleaver size of (a) $S=1$ (b) $S=n$, and (c) $S=s$.}
  \label{figure:schematic_bitinterleaver}
\end{figure}

We first consider the CP-MLC-ID without a bit-interleaver i.e., the interleaver size $S=1$, and using $d$ of 3, as shown in Fig. \ref{figure:schematic_bitinterleaver}(a). If the component decoder outputs an estimated component codeword $\hat{\zz}_{j'}$, the error $\ee_{j'}^\ss=\ee_{j'}^\zz$ propagates with a weight $|\ee_{j'}^\ss|=|\ee_{j'}^\zz|=\mu_{\min}$, which makes it difficult to correct the bit errors using the other component code.

Next, we consider CP-MLC-ID using a simple bit interleaver $(\S_1,\S_2,\S_3) = (\I,\S^S,\I)$ of sizes $S=n$ and $S=s$, as respectively shown in Fig. \ref{figure:schematic_bitinterleaver}(b) and (c). One bit of the codeword with the length of $n$ is XORed with $n/S$ bits for each other codewords in $\S^S$, where $\S^S=(\S^S)^{-1}$ holds. The error for each codeword $\ee^\zz_{j'}$ is reduced to $\ee_{j'}^{\ss}$ with $|\ee_{j'}^{\ss}|=\mu_{\min}/S$. Thus, even if the incorrectness $\bDelta_{j'}$ is large, the bit interleaver reduces the effect of error propagation by dispersing $\bDelta_{j}$.

Note that the above example of CP-MLC-ID over a BSC is shown to be intended to aid in understanding error propagation. However, further probabilistic behavior analysis is required for the optimal design of the bit interleaver. For example, since the iterative decoder of CP-MLC-ID relies on approximate belief propagation, the decoding performance may be influenced by short cycles in the factor graph of the component code and XOR operations. Similar to TPC families, another error floor may arise due to the design of the search for the candidates in the low-BER region\cite{9935118}. While a detailed evaluation of probabilistic behavior is important, it is beyond the scope of the current work, and we therefore leave it to future research.

\begin{figure}[!ht]
  \centering
  \begin{overpic}[scale=0.15]{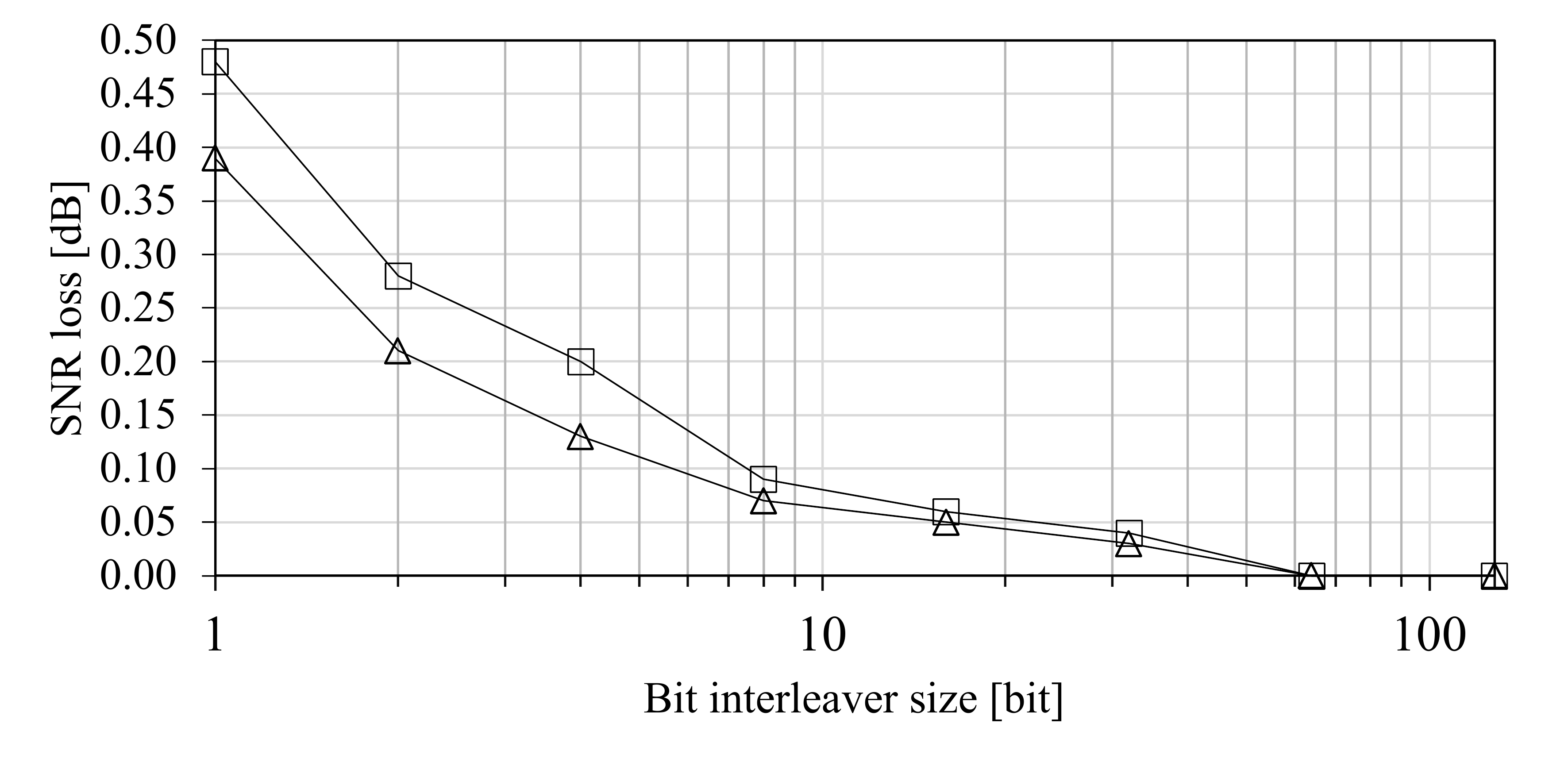}
  \end{overpic}  
  \caption{SNR loss with three and six iterations for CP-MLC-ID using $(128,106,8)$-eBCH code and OSD with $t=832$ candidate codewords for each bit interleaver size $S$. Squares and triangles represent three and six iterations, respectively. A total of 300 frames were transmitted.}
  \label{figure:interleaver_size_dependency}
\end{figure}

Figure \ref{figure:interleaver_size_dependency} shows the bit interleaver size dependency for CP-MLC-ID with three and six iterations using $(128,106,8)$-eBCH codes and OSD with 832 candidates. Note that the details of the code construction and the selection of candidates for OSD are described in Sec. \ref{section:sim_results} and Appendix \ref{appendix:candidatesOSD}, respectively. CP-MLC-ID with $S=1$ using three and six iterations has an SNR loss of 0.4 and 0.5 dB, respectively, compared to CP-MLC-ID with $S=128$. In contrast, the CP-MLC-ID with $S=8$ has an SNR loss of only about 0.1 dB. These findings demonstrate that CP-MLC-ID can improve the decoding performance by using only weak connections between codewords via XOR.

\begin{figure}[!ht]
  \centering
  \begin{overpic}[scale=0.32]{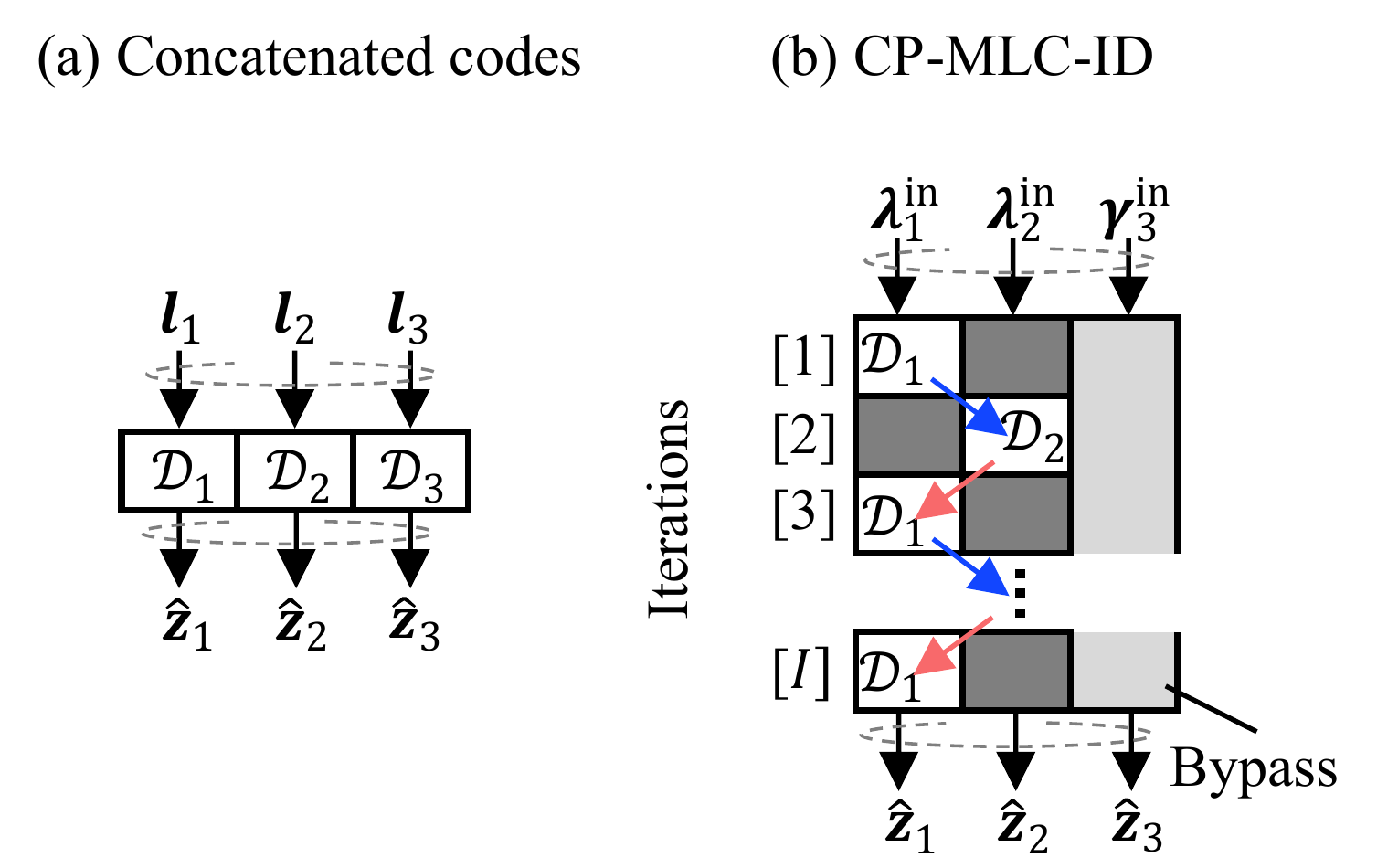}
  \end{overpic}  
  \caption{Schematic flow for (a) concatenated codes and (b) CP-MLC-ID with $d$ of 3.}
  \label{figure:iteration_flow}
\end{figure}

\begin{figure*}[!ht]
  \centering
  \begin{overpic}[scale=0.2]{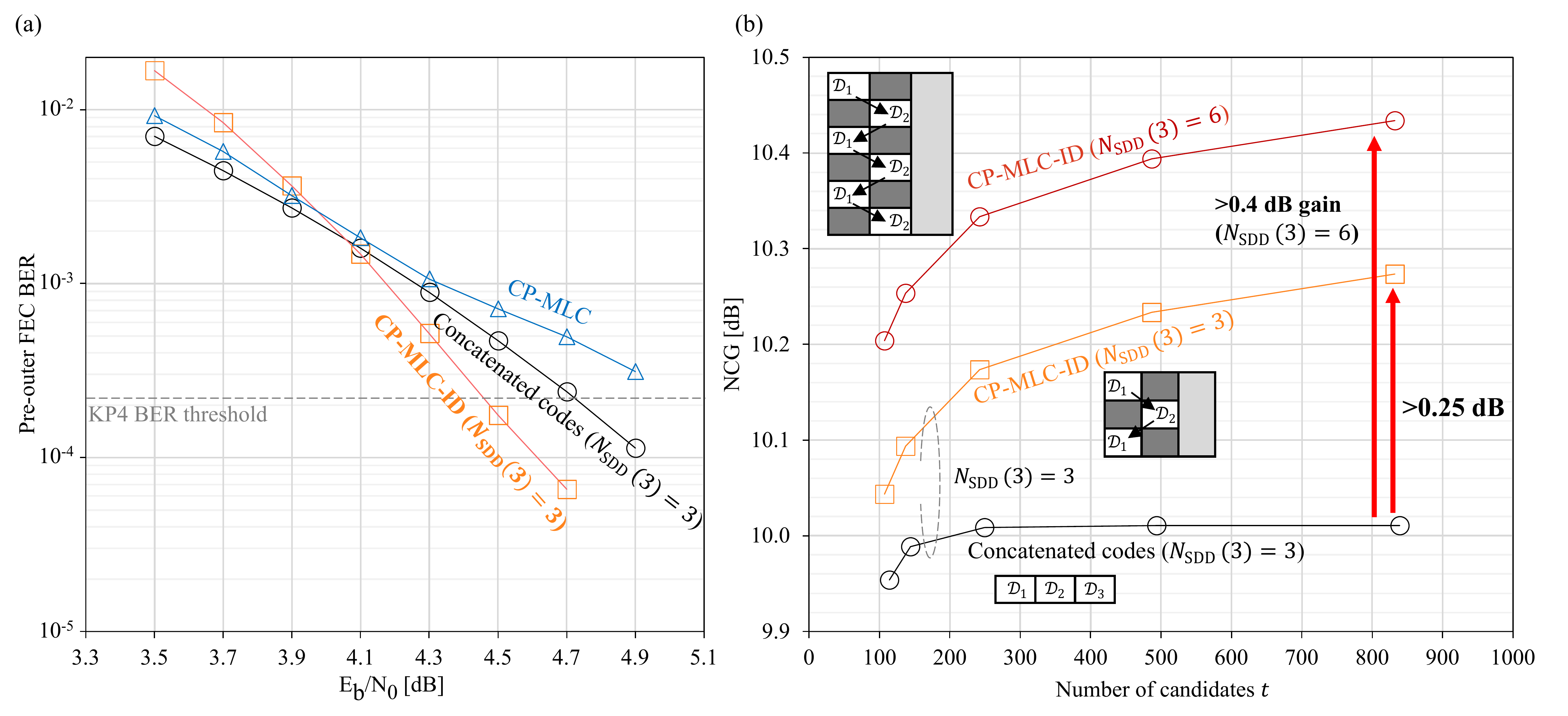}
  \end{overpic}
  \caption{(a) Decoding performance for concatenated code, CP-MLC, and CP-MLC-ID with number of SDDs $N_{\SDD}(3) = 3$. (b) NCG vs. number of candidates for concatenated codes, CP-MLC-ID with number of SDDs $N_{\SDD}(3) = 3$, and CP-MLC-ID with number of SDDs $N_{\SDD}(3) = 6$. Results were first evaluated in \cite{cpmlcid,cpmlchid}.}
  \label{figure:decoding_performance}
\end{figure*}

\subsection{Related work}
XOR operations are widely used in various code families to improve decoding performance. Similar to repeat accumulate (RA) codes \cite{divsalar1998coding,4155107}, the CP-MLC-ID copies, adds, and interleaves the bypassed bits into the component codewords. A key structural difference is that RA codes add bits sequentially using an accumulator, whereas CP-MLC-ID adds bypassed bits in parallel. The primary role of the XOR operation in CP-MLC-ID is to induce non-uniform channel capacities among the $d-1$ unreliable channels and one bypassed reliable channel. Owing to the parallel symmetry of the XOR structure, it may potentially achieve uniform capacity across the unreliable channels, in a manner similar to BICM-ID when iterative decoding is applied. This facilitates the design of component codes that correct errors in the unreliable channels. The auxiliary structure in CP-MLC-ID uses XOR only to extract extrinsic information, whereas accumulate-RA codes use XOR operations as part of the main decoding process.

Long block-length codes, such as non-binary low-density parity-check (LDPC) codes \cite{6336756}, TPC\cite{TPC_pyndiah}, and spatially coupled codes \cite{staircase,7152893}, consist of the component codes connected by XOR operations or shared bits. These codes improve decoding performance through iterative decoding, similar to CP-MLC-ID. However, CP-MLC-ID differs in that it connects the component codes indirectly and weakly via bypassed bits, which helps reduce bypassed-bit errors and provides extrinsic information to enhance the decoding performance of the component SDD.

\section{Numerical simulation}\label{section:sim_results}
This section describes the numerical simulation we conducted to evaluate the decoding performance.

\subsection{Code configuration}
We constructed about $19.5\%$–OH concatenated codes, CP-MLC, and CP-MLC-ID as shown in Table \ref{table:construction}. The decoding algorithm of the inner eBCH codes adopts the OSD. The outer code uses 5.56$\%$ KP4 code, which achieves the pre-FEC BER of $2.2\times 10^{-4}$ \cite{KP4_threshold}.

The concatenated codes, CP-MLC with $d$ of 2, and CP-MLC-ID with $d$ of 3 exploit the $(128,113,6)$-eBCH codes, $(128,113,10)$-eBCH codes, and $(128,113,8)$-eBCH codes, respectively. Note that the CP-MLC and CP-MLC-ID can use high-OH eBCH codes by bypassing the half- and one-third bits. The number of SDDs per three lanes, denoted by $N_{\SDD}(3)$, is equal to the number of iterations $I$, as shown in Fig. \ref{figure:iteration_flow}. CP-MLC-ID has the damping factors $\xi[i]$ as
\begin{align}
(\xi[1],\xi[2],\xi[3])=(0.3,1.0,1.0)
\end{align}
and
\begin{align}
(\xi[1],\xi[2],\xi[3],\xi[4],\xi[5],\xi[6])=(0.2,0.3,0.5,0.7,0.9,1.0),
\end{align}
by searching for suboptimal values in step of 0.1. Note that the optimization for each ${\xi[i]}$ depend on the error probability, the number of candidates of the inner decoder, the target bit error probability, and so on. CP-MLC-ID exploits the bit interleaver size of $S=n$ bits.

An 18.49$\%$-OH CP-MLC-ID with $d=3$ can be constructed using a $(256,215,12)$-eBCH and KP4 codes by doubling the block length to 256 while maintaining the same OH. However, a large number of candidate codewords are required for OSD of the component codes due to the large minimum distance \cite{OSD}. Moreover, increasing the block length significantly raises the computational complexity of \textit{Gaussian elimination}, which scales as $O(n^3)$, making OSD impractical for intra-DC applications.

\begin{table}[h]
  \caption{Code construction with outer KP4 codes}
  \label{table:construction}
  \centering
   \begin{tabular}{cccccccc}
    \hline
    Code &Total-OH\ $[\%]$ & Inner codes & OH\ $[\%]$\\
    \hline \hline
    Concatenated codes & 19.89 & $(128,113,6)$-eBCH & 13.27\\
    CP-MLC & 19.36 & $(128,99,10)$-eBCH & 29.29\\
    CP-MLC-ID & 19.53 & $(128,106,8)$-eBCH & 20.75\\
    \hline
   \end{tabular}
 \end{table}

\subsection{Numerical simulation results}
Figure \ref{figure:decoding_performance}(a) shows the decoding performance of the concatenated codes, CP-MLC, and CP-MLC-ID, which have $t=839$, $t=825$, and $t=832$ candidates, respectively. CP-MLC has an error floor caused by the reliable bit channel $V_{j}$, and therefore it cannot exceed the performance of concatenated codes under the KP4-BER threshold. In contrast, CP-MLC-ID outperformed the concatenated codes at the KP4 BER threshold because the inner $(128,106,8)$-eBCH codes correct a large part of the bit error and keep the error floor low enough by increasing the capacity of the reliable bit channel $U_i$, compared to the CP-MLC. Note that the error propagation effect of CP-MLC and CP-MLC-ID degrades the BER in high BER regions.

Figure \ref{figure:decoding_performance}(b) shows the NCG for concatenated codes, CP-MLC-ID with $N_{\SDD}(3)=3$, and $N_{\SDD}(3)=6$, for each candidate. The NCG of the concatenated codes is saturated as it approaches the MLD performance. The CP-MLC-ID with $N_\SDD = 3$ can improve the decoding performance by 0.25 dB or more compared to the concatenated codes. The CP-MLC-ID with $N_{\SDD}(3)=6$ can improve the NCG by 0.4 dB or more by doubling the number of SDDs $N_{\SDD}(3)=6$.

The decoding complexity of CP-MLC-ID consists of the updating $\blambda^\IN_j$ (as shown in Eq. \eqref{equation:lambda}), and the bit interleaver, in addition to the SDDs. Equation \eqref{equation:lambda} requires processing on the order of $O(n)$ for the boxplus operator. Thus, the updating cannot be ignored in the evaluation of complexity with Eq. \eqref{equation:lambda} when using inner codes with a small number of candidates. The bit interleaver also has the potential to become a major contributor to complexity because it requires the shuffling LLR, which has 3- to 5-bit quantization. Therefore, the size of the bit interleaver must be carefully designed. Evaluation of complexity and power consumption is difficult without a well-designed and efficient circuit design, which will be the subject of future work.

\section{Conclusion}\label{section:conclusion}
In this paper, we showed that channel-polarized multilevel coding with iterative decoding (CP-MLC-ID) improves the decoding performance under low-complexity SDD. The CP-MLC-ID iteratively corrects large errors by using the high-OH inner codes with SDD in the unreliable bit channels, bypassing the SDD in a highly reliable bit channel. CP-MLC-ID outperforms the maximum NCG by up to 0.25 and 0.4 dB for the same number of SDDs and twice the number of SDDs, respectively, compared to concatenated codes, by using only XOR between the inner codes.

\appendices
\section{Factor graph representation of CP-MLC-ID\cite{cpmlchid}}\label{appendix:factorgraph}
\begin{figure}[!ht]
  \centering
  \begin{overpic}[scale=.25]{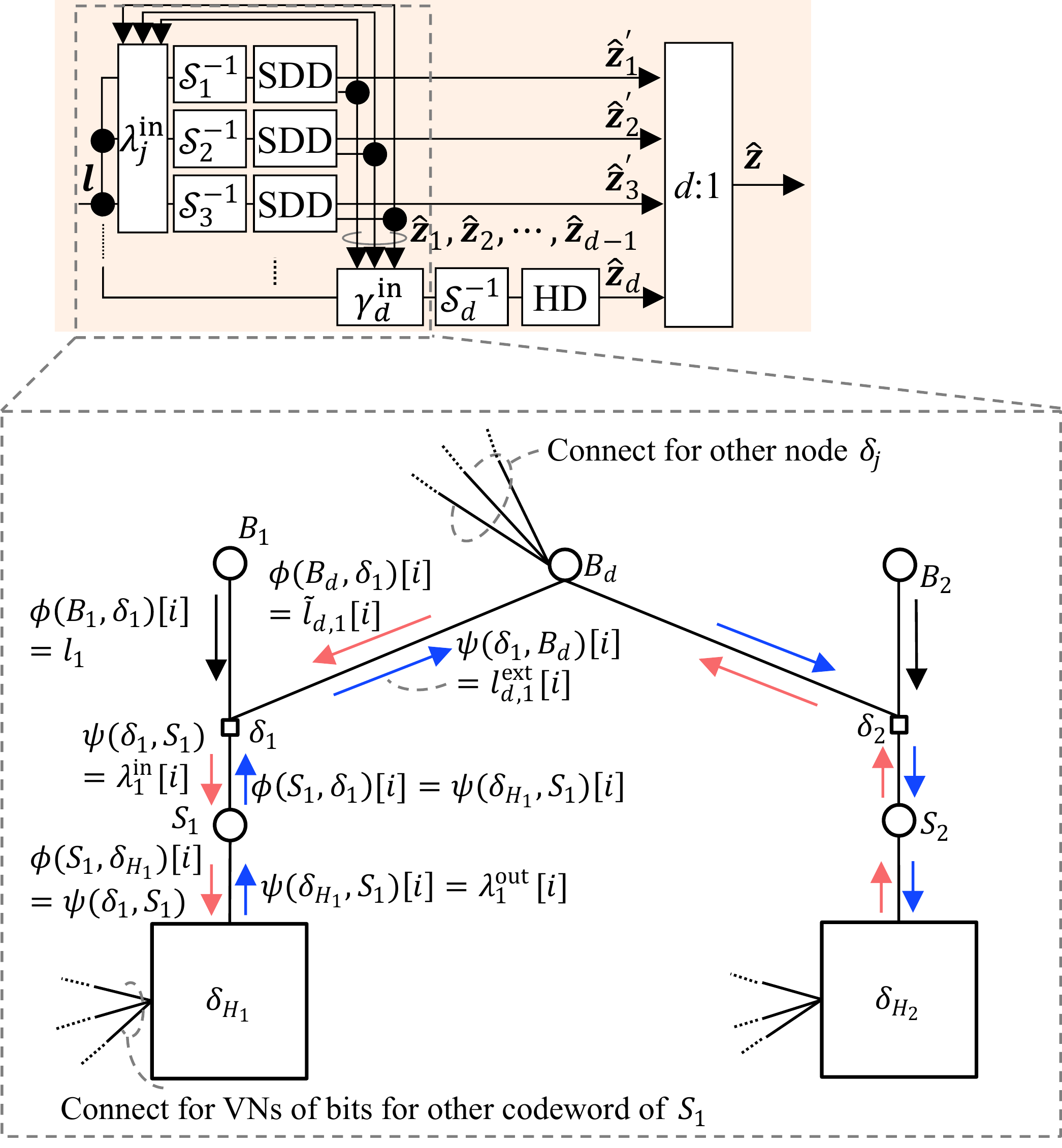}
  \end{overpic}
  \caption{Factor representation of decoder for CP-MLC-ID.}
  \label{figure:factorgraph}
\end{figure}

To qualitatively assess the decoder's behavior, we give the factor-graph $\G(\V,\C)$ \cite{moderncodingtheory} of the algorithm in Fig. \ref{figure:factorgraph}. We consider the \textit{variable node} (VN)  $\V = \{B_1,B_2,\cdots,B_d, S_1,S_2,\cdots,S_{d-1}\}$ and \textit{check node} (CN) $\C = \{\delta_1,\delta_2,\cdots,\delta_{d-1},\delta_{H_1},\delta_{H_2},\cdots,\delta_{H_{d-1}}\}$, where each $\delta_j$ represents an XOR constraint and each $\delta_{H_j}$ represents to a constraint derived from the parity-check matrix of the component codes.

The message scheduling of CP-MLC-ID is given by
\begin{align}
&\phi(S_1,\delta_{H_1})[1]\to\psi(\delta_{H_1},S_1)[1]\to\phi(S_1,\delta_1)[1]\notag\to\psi(\delta_1,B_d)[1]\\
&\to\phi(B_d,\delta_2)[2]\to\psi(\delta_2,S_2)[2]\to\cdots\to\psi(\delta_{j},B_d)[I_{\text{max}}],\label{equation:HID}
\end{align}
with $j = (i-1\text{\ mod } d-1) + 1$, where $\phi$ and $\psi$ are the messages of VN to CN and vice versa, respectively. CP-MLC-ID then calculates the estimated bypass bits $\hat{\zz}_d$ from Eq. \eqref{equation:reliable_bits}. 

\section{Generation rule for candidates in OSD}\label{appendix:candidatesOSD}
In this section, we explain how to select codeword candidates for OSD. OSD can order the decoding metric (e.g., LLR $\ll$) and then convert the code space $C(G)$ by generating the generator matrix $G$ into a new code space $C(G')$ where the parity bits correspond to the positions of $n-k$ linearly independent LRBs. The candidate codeword set $\Z$ is obtained by $\zz = (\ii\oplus \tt)G'$ for each $\tt\in\T$, which is a \textit{flipping set}. Finally, the OSD outputs the most likely codeword $\hat{\zz}$, which is the minimum Euclidean distance between the decoding metric and $\zz\in\Z$.

The flipping set as \textit{order-$m$} $\T_m$ consists of flipping $m$-bit or fewer. We define the flipping set as \textit{semi-order}-$m$ $\T_2(m_1,m_2)$, which selects two bits in the range of $0,1,2,\cdots m_1$ and $0,1,2,\cdots m_2$, respectively. The number of candidates $|\T_2(m_1,m_2)|$ is given by
\begin{align}
|\T_2(m_1,m_2)| = {}_{m_1}C_2 - {}_{m_1-m_2}C_2\ad{.}
\end{align}
In the candidate selection discussed in Sec. \ref{section:sim_results}, we choose the parameters as shown in Tab. \ref{table:osd_paramaters}. Note that the parameters are suboptimal, which means that we can improve the performance by optimizing the parameter.

\begin{table}[h]
  \caption{Candidates for OSD}
  \label{table:osd_paramaters}
  \centering
   \begin{tabular}{cccccccc}
    \hline
    Code &\vline\vline & Candidate set $\T$ & \vline\vline&$|\T|$\\
    \hline \hline
    (128,113,6)-eBCH codes & \vline\vline&  $\T_0\cup\T_1$ & \vline\vline& 114\\
    & \vline\vline& $\T_0\cup\T_1\cup\T_2(10,4)$& \vline\vline& 144\\
    & \vline\vline& $\T_0\cup\T_1\cup\T_2(20,9)$&\vline\vline& 249\\
    & \vline\vline& $\T_0\cup\T_1\cup\T_2(30,19)$& \vline\vline&494\\
    & \vline\vline& $\T_0\cup\T_1\cup\T_2(40,29)$&\vline\vline& 839\\
    \hline
    (128,106,8)-eBCH codes& \vline\vline& $\T_0\cup\T_1$& \vline\vline&107\\
    & \vline\vline& $\T_0\cup\T_1\cup\T_2(10,4)$& \vline\vline&137\\
    & \vline\vline& $\T_0\cup\T_1\cup\T_2(20,9)$& \vline\vline&242\\
    & \vline\vline& $\T_0\cup\T_1\cup\T_2(30,19)$& \vline\vline&487\\
    & \vline\vline& $\T_0\cup\T_1\cup\T_2(40,29)$& \vline\vline&832\\
    \hline
    (128,99,10)-eBCH codes &\vline\vline& $\T_0\cup\T_1\cup\T_2(40,29)$& \vline\vline&825 \\
    \hline
   \end{tabular}
 \end{table}

\ifCLASSOPTIONcaptionsoff
  \newpage
\fi

\bibliographystyle{IEEEtran}
\bibliography{bare_jrnl.bib}

\end{document}